\documentclass[aps,prB,superscriptaddress,twocolumn]{revtex4}
\usepackage{amsmath}
\usepackage{amsfonts}
\usepackage{graphicx}
\usepackage{color}
\usepackage{dsfont}
\usepackage{verbatim}
\usepackage{mathtools}
\usepackage[normalem]{ulem}

\begin{document}

\newcommand{\beq}{\begin{equation}}
\newcommand{\eeq}{\end{equation}}
\newcommand{\beqa}{\begin{eqnarray}}
\newcommand{\eeqa}{\end{eqnarray}}
\newcommand{\ben}{\begin{enumerate}}
\newcommand{\een}{\end{enumerate}}
\newcommand{\hs}{\hspace{1.5mm}}
\newcommand{\vs}{\vspace{0.5cm}}
\newcommand{\note}[1]{{\color{red}  #1}}
\newcommand{\notea}[1]{{\bf #1}}
\newcommand{\new}[1]{{#1}}
\newcommand{\ket}[1]{|#1 \rangle}
\newcommand{\bra}[1]{\langle #1|}
\newcommand{\im}{\dot{\imath}}
\newcommand{\tg}[1]{\textcolor{blue}{#1}}

\title{Quantum Entanglement of the Sachdev-Ye-Kitaev Models}

\author{Chunxiao Liu}
\affiliation{Department of Physics, University of California, Santa Barbara, CA, 93106-9530}
\author{Xiao Chen}
\affiliation{Kavli Institute for Theoretical Physics CA 93106-4030, USA}
\author{Leon Balents}
\affiliation{Kavli Institute for Theoretical Physics, University of California, Santa Barbara, CA 93106-4030, USA}

\date{\today}

\begin{abstract}
The Sachdev-Ye-Kitaev (SYK) model is a
quantum mechanical model of fermions interacting with $q$-body random couplings.  For $q=2$, it describes free particles, and is non-chaotic in the many-body sense, while for $q>2$ it is strongly interacting and exhibits many-body chaos. In this work we study the entanglement entropy (EE) of the SYK$q$ models, for a bipartition of $N$ real or complex fermions into subsystems containing $2m$ real/$m$ complex fermions and $N-2m$/$N-m$ fermions in the remainder. For the free model SYK$2$, we obtain an analytic expression for the EE, derived from the $\beta$-Jacobi random matrix ensemble.   Furthermore, we use the replica trick and path integral formalism to show that the EE is {\em maximal} for when one subsystem is small, i.e. $m\ll N$, for {\em arbitrary} $q$.  We also demonstrate that the EE for the SYK4 model is noticeably smaller than the Page value when the two subsystems are comparable in size, i.e. $m/N$ is $O(1)$.  Finally, we explore the EE for a model with both SYK2 and SYK4 interaction and find a crossover from SYK2 (low temperature) to SYK4 (high temperature) behavior as we vary energy. 
\end{abstract}

\maketitle

\section{Introduction}

The Sachdev-Ye-Kitaev (SYK) models \cite{SachdevYe1993, Sachdev2015, Kitaev15talk, Madacena16Remarks,2016JHEP...04..001P} are large-$N$ solvable zero-dimensional systems of $N$ Majorana or complex fermions coupled through all-to-all $q$-fermion interactions.  They have drawn considerable interest in high energy physics for their emergent conformal symmetry and holographic duality, and excite the condensed matter community as solvable examples of strongly interacting quantum critical non-Fermi liquids, which lack any quasiparticle description \cite{Banerjee16Solvable}.   This is reflected in the solution of the Green's function, the scaling dimension of Majorana fields, and the non-zero entropy at zero temperature \cite{Madacena16Remarks}. At $q>2$, the model is also maximally chaotic which can be seen from its quantum Lyapunov exponent defined from the out-of-time-order correlator \cite{Larkin1969, Shenker2014, Kitaev15talk,Madacena16Remarks, 2016JHEP...04..001P}. Starting from the original SYK4 model, many extensions have been proposed and studied which retain the interacting and chaotic nature \cite{Banerjee16Solvable, Bi17instability, YizhuangYou17SYK, Gu2017,jian2017model,gu2017energy, song2017strongly, chen2017competition, jian2017solvable, 2017arXiv170702197G}. On the other hand, the $q=2$ case is a free fermion model that can be completely solved by random matrix theory. It realizes a zero dimensional disordered fermi liquid in the large $N$ limit, and both many-body chaos and zero temperature entropy are wholly absent. Given these facts, it is natural to study the phase transition between these two regimes \cite{Banerjee16Solvable,Bi17instability,song2017strongly}.

In this paper we study the entanglement entropy (EE) of the SYK$q$ model and discuss the differences between the SYK2 and SYK$q$ models with $q>2$ from the entanglement perspective.  EE, a concept borrowed from quantum information theory, can quantify the quantum entanglement of different subsystems. The von Neumann EE for subsystem $A$ is concisely defined as $S_A = -\mathrm{Tr}[\rho_A\ln \rho_A]$, where $\rho_A$ is the reduced density matrix of $A$. For the SYK2 model, we show that the EE can be analytically computed in closed form by connecting to the concept of Jacobi ensemble in random matrix theory \cite{loggasrandommatrices}. For SYK$q$ with $q>2$, since the model is maximally chaotic in the large $N$ limit, one would naturally presume that it is also maximally entangled \cite{WenboFu16Numerical}.   We numerically calculate EE and find that in the limit when $A$ is much smaller than the total system size, this is correct: the ground state EE for the SYK4 is maximally entangled.  However, this is also true for SYK2, which is not many-body chaotic at all.  We then show that these observations can be explained analytically by using the standard replica trick in the path integral formalism.   Deviations from maximal entanglement are seen when the two subsystems obtained from bipartition are comparable in size: we find that the EE for the SYK4 model is much larger than that for the SYK2 model,  yet both are still smaller than the Page value, which can be considered as the EE for the Gaussian unitary random matrix ensemble (GUE). Further $1/N$ extrapolation indicates that the difference between the EE of the SYK4 and Page value persists to the thermodynamic limit, showing that the SYK4 model is not maximally entangled.

\begin{figure}[t]
\centering
\includegraphics[width=0.5\textwidth]{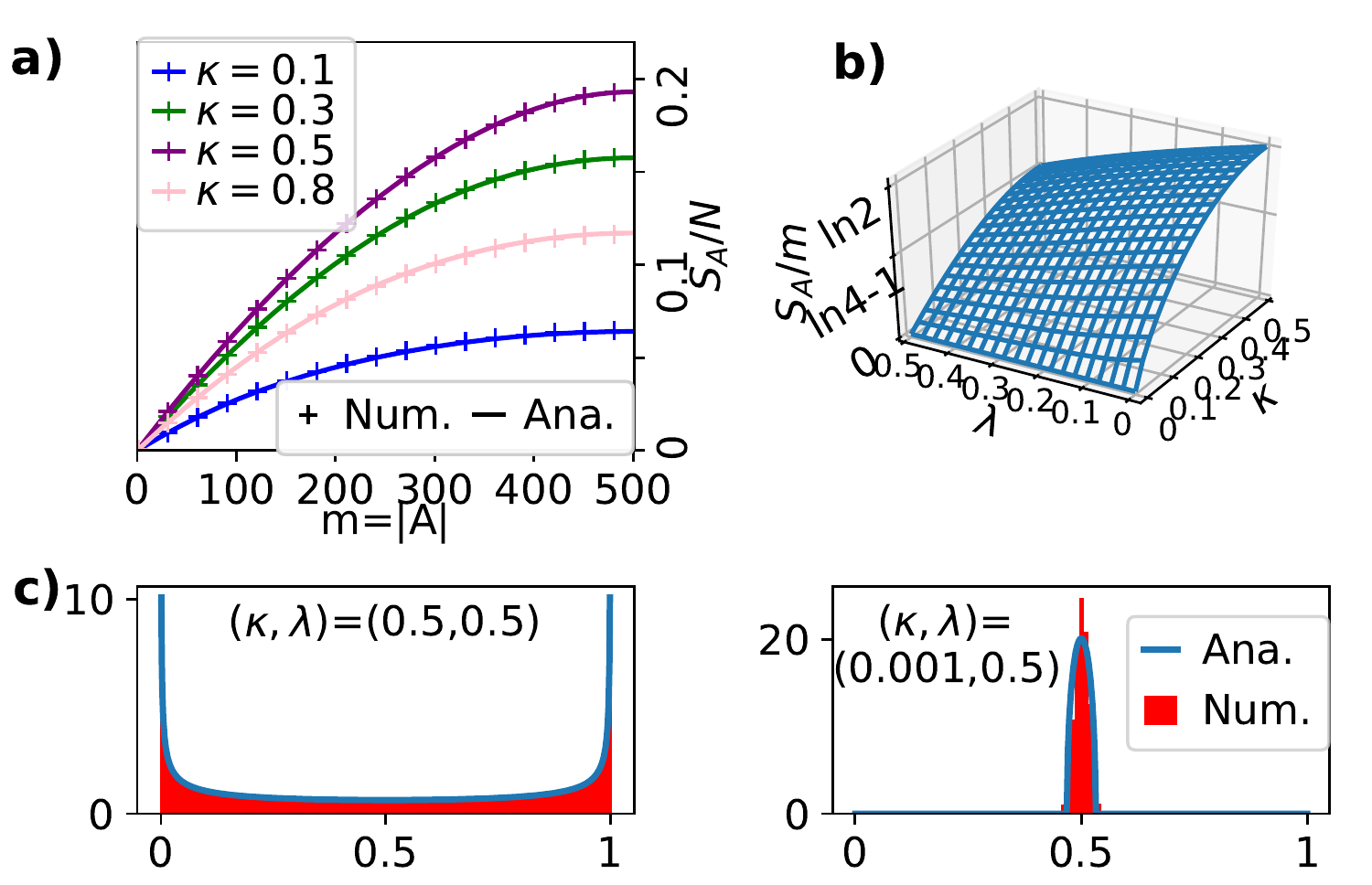}
\caption{a). Scaling of EE of complex SYK2 model with subsystem size at different filling factors, numerical versus analytical result, with fixed $N=1000$. b). EE per site as function of $\kappa$ and $\lambda$. c). The Wachter Law [Eq.~\eqref{wachter}] of the complex SYK2 model, numerical (red) versus analytical (blue) result.}\label{Fig_1}
\end{figure}

We further study a model which includes both SYK2 and SYK4 interactions. At low temperature,  the physics is dominated by the SYK2 term, which may be understood from the relevance, in the scaling sense, of the SYK2 interaction at the SYK4 fixed point.\cite{Banerjee16Solvable,Bi17instability,song2017strongly} This is consistent with our ground state EE result which shows that EE decreases as we increase $N$ and approaches the SYK2 value in the thermodynamical limit. For a highly excited state, we find that the EE is the same as that for SYK4 model and Page value, indicating there is crossover from low energy SYK2 physics to high energy SYK4 physics.

The rest of the paper is organized as follows: In Sec.~\ref{sec:model}, we briefly review SYK model and discuss the methods to compute EE. In Sec.~\ref{sec:syk2}, we analytically compute the EE of SYK2 model and analyze the scaling behavior of EE in various limits. In Sec.~\ref{sec:syk4}, we numerically study EE of SYK2+SYK4 model and explore the crossover behavior from SYK2 to SYK4 physics. We further discuss the replica trick method in Sec.~\ref{sec:replica}. We conclude in Sec.~\ref{sec:conclusion} with some final remarks.

\section{Models and Methods}
\label{sec:model}
The Hamiltonians of the SYK$q$ models at $q=2$ and $4$, both complex and Majorana fermion versions, are written as
\begin{subequations}
\label{SYK_model}
\begin{eqnarray}
H_{2,c} &=& \sum\limits_{i,j=1}^N T_{ij} c^\dag_i c_j\label{SYK2c}-\mu\sum_{i=1}^Nc_i^{\dag}c_i,\\
H_{2,\chi}&=& \sum\limits_{i,j=1}^N iJ_{ij} \chi_i \chi_j,\label{SYK2m}\\
H_{4,c} &=& \sum\limits_{\{i,j,k,l\}} T_{ij;kl} \,c^\dag_i c^\dag_j c_k c_k\label{SYK4c}-\mu\sum_{i=1}^Nc_i^{\dag}c_i,\\
H_{4,\chi} &=& \sum\limits_{1\leq i<j<k<l\leq N} J_{ijkl} \,\chi_i\chi_j\chi_k\chi_l.
\label{SYK4m}
\end{eqnarray}
\end{subequations}
The matrix $T=(T_{ij})$ is hermitian and $J=(J_{ij})$ is antisymmetric, with all Gaussian entries with zero mean and variance $\frac{\mathcal{J}^2}{N}$, where $\mathcal{J}$ is some constant; the tensor $T_{ij;kl}$ is complex with appropriate symmetry to ensure hermiticity and the tensor $J_{ijkl}$ is real, both with Gaussian entries of zero mean and variance $\frac{3!\mathcal{J}^2}{N^3}$ \cite{Madacena16Remarks,WenboFu16Numerical}. For the complex SYK$q$ model, a chemical potential $\mu$ is included to tune the filling factor.

We  study the EE for the models in Eq.~\eqref{SYK_model}. As SYK$q$ are zero-dimensional models, the bipartition is made by choosing subsystem $A$ to consist of $2m$ random Majorana or $m$ random complex fermions, with the complement called $B$. Since SYK$q$ includes all to all coupling between fermions, we expect that the disorder averaged EE for subsystem A has the form $\overline{S_A}=\alpha m-\gamma$, describing a ``volume law'' for EE.   There is no regime of ``area law'' due to the complete non-locality of interactions.  The main purpose of this paper is to investigate the coefficient $\alpha$ and the possible subleading correction $\gamma$.   Below we will resort to both analytical methods, which involve random matrix theory and the path integral formalism, and numerical means, mainly exact diagonalization, to study the EE.

\section{Analytical results for SYK2}
\label{sec:syk2}
The complex fermion SYK2 model (Eq.~\eqref{SYK2c}) is a free Hamiltonian in which $T$ is a hermitian random matrix belonging to the Gaussian unitary ensemble (GUE) \cite{loggasrandommatrices}. The eigenvalue distribution of GUE satisfies the ``semi-circle'' law, and the diagonalizing matrix $U$, composed of column eigenvectors, is uniformly distributed in the space of unitary matrices according to the Haar measure.

For a free fermion model, the EE of a subsystem $A$ can be directly calculated from the reduced two point correlation function matrix $C_A$ with each entry $\left(C_A\right)_{ij} = \langle c^\dag_i c_j\rangle$ for $i,j \in A$ \cite{Vidal03Entanglement, Peschel02Reduced}. A key observation is that $C_A= V^\dag V$ belongs to the Jacobi random matrix ensemble at $\beta=2$ \cite{loggasrandommatrices} (see Appendix \ref{sec: Jacobi}), where $V$ is the $k$-by-$m$ upper-left block of $U^T$, and $k$, $m$ are the filling number of energy bands and subsystem size of $A$, respectively. The distribution of the eigenvalue $x$ of $C_A$ in the large $N$ limit can be derived from the joint probability distribution of the Jacobi ensemble and satisfies the Wachter law \cite{wachter1980law,loggasrandommatrices},
\begin{equation}\label{wachter}
\begin{aligned}
f(x,\kappa, \lambda) &= \frac{1}{2\pi\lambda} \frac{\sqrt{(\lambda_+-x)(x-\lambda_-)}}{x(1-x)}1_{[\lambda_-,\lambda_+]}\\
&+\left(1-{\kappa}/{\lambda}\right)\Theta(\lambda-\kappa)\delta(x),\qquad
\lambda,\kappa \in (0,1/2],
\end{aligned}
\end{equation}
where $\lambda_\pm = (\sqrt{\kappa(1-\lambda)}\pm \sqrt{\lambda(1-\kappa)})^2$, $\lambda = \frac{m}{N}$ and $\kappa = \frac{k}{N}$.
The $\delta$ function exists when $V^\dag V$ is singular. The continuum part of $f(x,\kappa, \lambda)$, shown in Fig.~\ref{Fig_1}c), is nonzero only on the support ${[\lambda_-,\lambda_+]}$. The Wachter law, Eq.~\eqref{wachter}, does not depend on the value of $\beta$, and holds also for orthogonal ($\beta=1$) and quaternion matrices ($\beta=4$) -- see Appendix \ref{sec: Jacobi}. 
The EE is therefore
\begin{equation}\label{EE1}
\begin{aligned}
\overline{S_A(\kappa,\lambda)} = m\int  \left[-x \ln x -(1-x) \ln(1-x)\right]
f(x,\kappa,\lambda) dx.
\end{aligned}
\end{equation}
Due to particle-hole symmetry, the EE of the Majorana SYK2 model has the same scaling behavior as that of the complex fermion SYK2 model at $\kappa=1/2$ (Appendix \ref{sec:Maj}).

\begin{figure}
\centering
\includegraphics[width=0.5\textwidth]{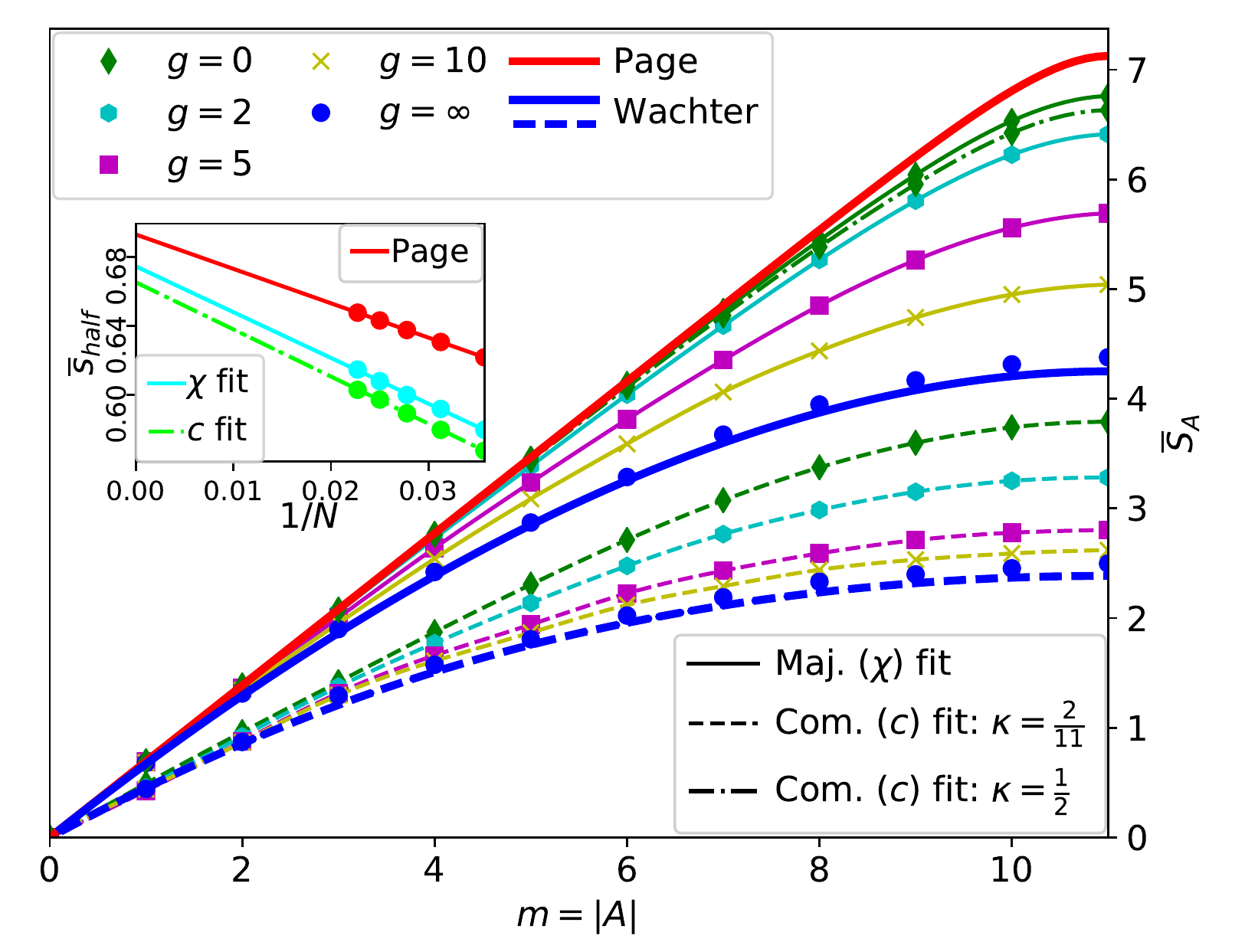}
\caption{EE scaling of the ground state of both the Majorana fermion ($\chi$) and the complex fermion ($c$) SYK4+SYK2 models, averaged over 10 samples. $m=|A|$ is the subsystem size of complex SYK, while for Majorana SYK the subsystem size is $2m$. The fermion numbers are $N=22$ for $c$ and  $N=44$ for $\chi$. The Hamiltonian is $H = H_{4,c\slash\chi}+\frac{g}{N} H_{2,c\slash\chi}$. The complex fermion model has fixed particle number $k=4$ (dashed lines) or $k=11$ (dash-dotted lines). Page and Wachter scalings are plotted according to Eq.~\eqref{Page} and Eq.~\eqref{EE1}. The inset shows finite size scaling of the EE at subsystem/system=1/2.}\label{SYK_N_44}
\end{figure}

We verify the theoretical result for the EE in Eq.~\eqref{EE1} by comparing it with a numerical calculation in a large system with $N=1000$ [Fig.~\ref{Fig_1}a)]. In fact the function $f(x,\lambda,\kappa)$ has several limiting forms for which EE can be computed analytically. For instance, when the subsystem is much smaller than the total system, $f$ approaches the $\delta$ distribution [Fig.~\ref{Fig_1}c) right] and the EE is \begin{equation}\label{SAlambda0}
\overline{S_A} \xrightarrow{\lambda\rightarrow 0}
m\left(-\kappa \ln \kappa - (1-\kappa)\ln (1-\kappa)\right).
\end{equation}
This result has also been obtained in Ref. \onlinecite{Magan16randomfree}. In particular, when $\kappa=1/2$, i.e. complex fermion at half-filling or Majorana fermion, the EE is simply equal to $m\ln 2$. This result agrees with simply counting the degrees of freedom  of $A$ and exhibits maximal entanglement between $A$ and $B$. However when $A$ is comparable to $B$ in size, $f$ disperses in the whole $[0,1]$ interval [Fig.~\ref{Fig_1}c) left] and the entanglement deviates from the maximal value; in particular, when $\lambda=1/2$ we have $S_{A}= m(2 \ln 2- 1)$. This result is consistent with the bounds given for the quadratic fermionic Hamiltonian \cite{Vidmar2017}. 

Eq.~\eqref{EE1} features a symmetry $(\lambda,m)\leftrightarrow (\kappa,k)$ apart from the trivial symmetries $\lambda\leftrightarrow 1-\lambda$ and $\kappa \leftrightarrow 1-\kappa$. Therefore in the zero filling factor limit, $\overline{S_A} \xrightarrow{\kappa\rightarrow 0}
k\left(-\lambda \ln \lambda - (1-\lambda)\ln (1-\lambda)\right)$ by analogy with Eq. \eqref{SAlambda0}. The quantity entanglement entropy density, defined as $S_A/m$, measures the average EE for each fermion and is plotted in Fig. \ref{Fig_1}b). Clearly the maximal value is reached only when $(\lambda,\kappa)=(0,1/2)$, and the EE density is a descending function of both $\lambda$ and $\frac{1}{2}-\kappa$. 

\section{Numerical and analytical results for SYK4+SYK2}
\label{sec:syk4}
The complex SYK4 model at half filling has been investigated in a small system in Ref.\ \onlinecite{WenboFu16Numerical}, where the authors reported that this model exhibits maximal entanglement. Here via sparse matrix diagonalization we numerically calculate the ground state EE of both the complex fermion SYK4 for $N$ up to $N=22$ and the Majorana fermion versions for $N$ up to $N=44$, where for the former we choose two filling factors. The results are shown in Fig.~\ref{SYK_N_44}. We see that when $m\ll N/2$, the EE density $\overline{S_A}/m$ is equal to
$-\kappa \ln \kappa-(1-\kappa)\ln (1-\kappa)$. This is the same as Eq.~\eqref{SAlambda0} for SYK2 and in fact is true for all SYK$q$ models (see below). As we increase $m$, the EE for SYK4 is much larger than that for SYK2. However, from the results of $N=44$, it is hard to predict the analytical form in the thermodynamical limit.

To better understand the scaling of EE around $\lambda=1/2$ in the thermodynamical limit, we conduct  finite size scaling analysis (see inset of Fig.~\ref{SYK_N_44}) and compare EE of both Majorana and complex fermion SYK4 models at $\kappa=1/2$ with Page value \cite{Page93entropy}
\begin{equation}\label{Page}
 \overline{S_{A,P}} = m \ln 2 - \frac{|\mathcal{H}_A|}{2|\mathcal{H}_B|},
\end{equation}
where $|\mathcal{H}_A|$ and $|\mathcal{H}_B|$ are the dimensions of the Hilbert space of $A$ and $B$ respectively. The Page value is the EE for a random pure state (eigenstate of GUE) and is equal to maximal EE up to a small constant, which is $1/2$ when $|\mathcal{H}_A|=|\mathcal{H}_B|$. Fig.~\ref{SYK_N_44} shows that the Page value (red line) and SYK4 (green solid and dashed lines) are the same when $m\ll N/2$ but differ around $\lambda=1/2$. One naturally asks how much this lowering in SYK4 is due to finite size effects. To address this question, we further check the EE density of SYK4 at $\lambda=1/2$ for different $N$, denoted $\overline{s}_{\text{half}}$, and find excellent linearity between $\overline{s}_{\text{half}}$ and $1/N$ (see the inset of Fig.~\ref{SYK_N_44}). This suggests that at $\lambda=1/2$, $\overline{S_A}=m\overline{s}_{\text{half}} -\gamma$ with $\overline{s}_{\text{half}}=0.675(0.665)$ for Majorana (complex) fermions obtained from $1/N$ extrapolation. $\overline{s}_{\text{half}}$ is smaller than $\ln 2$, suggesting the deviation from Page state in the thermodynamical limit. On the other hand, $\gamma=0.666\ (0.686)$ and is larger than $1/2$ in the Page value. Both these results demonstrate that the SYK4 ground state is less entangled than Page state around $\lambda=1/2$. However, the difference should disappear as we increase $q$ since the Hamiltonian becomes less sparse and approaches a member of the GUE.

We also remark that the thermal entropy at zero temperature is smaller than the EE in the SYK4 model. 
It is known that Majorana SYK$q$ model (with $q\geq 4$) has zero temperature residual entropy in the large N limit with this particular form  
$\frac{\overline{S_0}}{N/2} \sim \ln 2 - f(q)$, where $f(q)$ is some function of $q$
\cite{Kitaev15talk, Madacena16Remarks}. When $q=4$, this value is around $0.4648$ and is much smaller than the ground state EE density. This difference disappears as we increase $q$.

We further numerically calculate the ground state EE of $H=H_{4,c/\chi}+\frac{g}{N}H_{2,c/\chi}$ for different couplings $g$, see Fig.~\ref{SYK_N_44}. It is known that the SYK2 interaction is a relevant perturbation which induces a flow from SYK4 (a non-fermi liquid) behavior to a free fixed point \cite{Banerjee16Solvable,Bi17instability,2017arXiv170702197G}. This physics is also reflected in the scaling of EE at finite $N$, in two aspects: (i) for a fixed $N$, the EE is lowered as $g$ is increased, which can be clearly observed in Fig.~\ref{SYK_N_44}, (ii) for a fixed $g$, as we increase $N$, EE for the same subsystem size remains the same. This is consistent with the scaling dimension of fermion operator equal to $1/2$. 

\begin{figure}[t]
\centering
\includegraphics[width=0.45\textwidth]{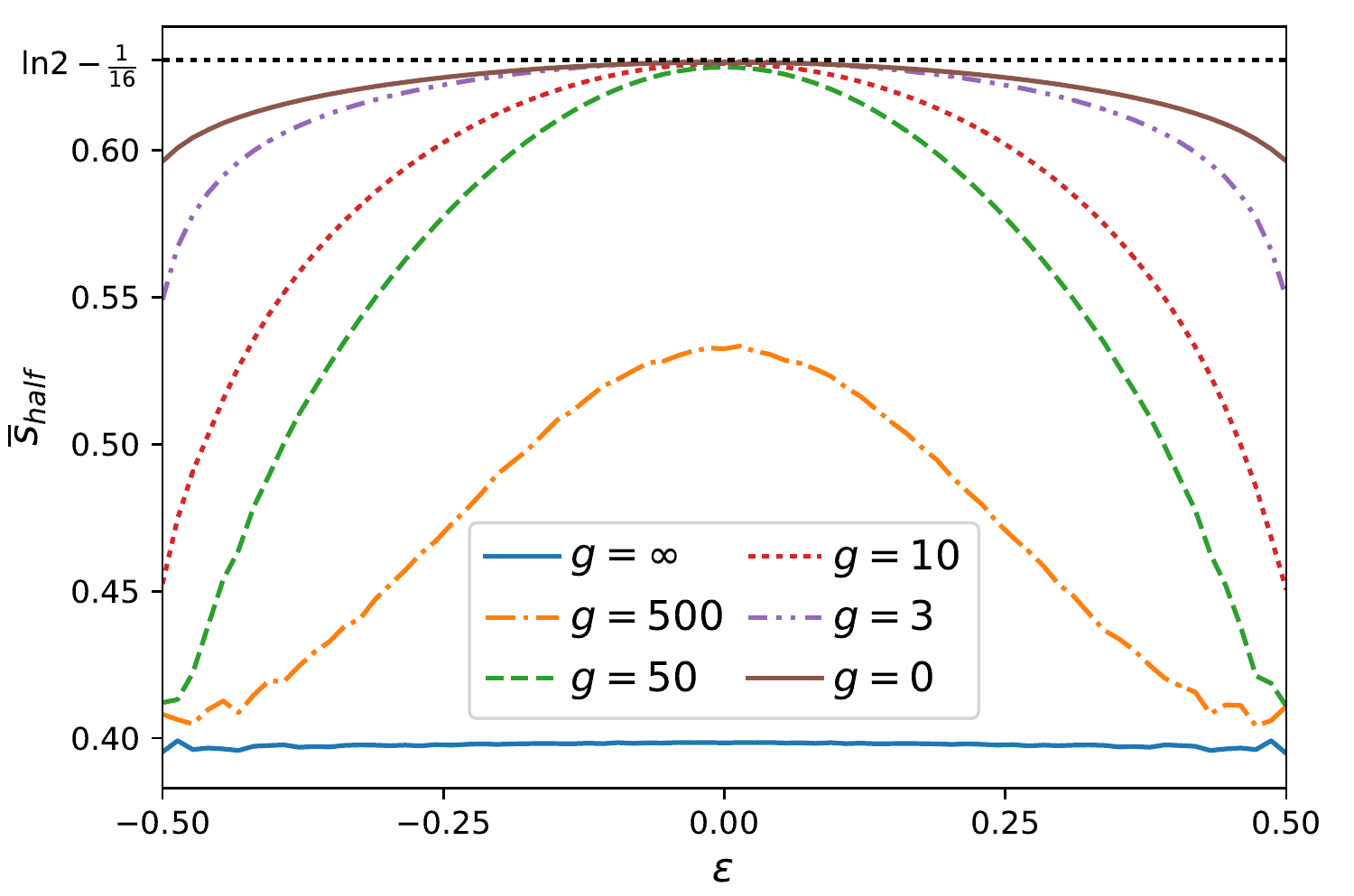}
\caption{EE as a function of energy for the Majorana SYK4+SYK2 models with different $g$, averaged over 10 samples. The energy has been normalized to the interval $[-0.5,0.5]$. We fix subsystem/system=1/2 and $N=32$. The flat dotted line (black) on the top is the Page value.}\label{SYK_N_32_fnt}
\end{figure}

On the other hand, as we increase the energy and move to a highly excited state, there is a ``crossover" from SYK2 physics to SYK4 physics and we expect that the EE for the highly excited state should be the same as that for the pure SYK4 interaction.\cite{song2017strongly} To verify this statement, we first study the excited state EE for the pure SYK4 and the SYK2 models and show that they are different (see Fig.~\ref{SYK_N_32_fnt}). We numerically find that the EE of a highly excited state for the SYK4 model is the same as Page value. In contrast, for the SYK2 model, both excited and ground state EE exhibit the same scaling since $C_A$ belongs to the Jacobi ensemble. As we introduce SYK4 interactions to the SYK2 model, the EE for the highly excited state drastically changes and becomes the Page value, the same as that for the pure SYK4 model.

\begin{figure}[t]
\includegraphics[width=0.4\textwidth]{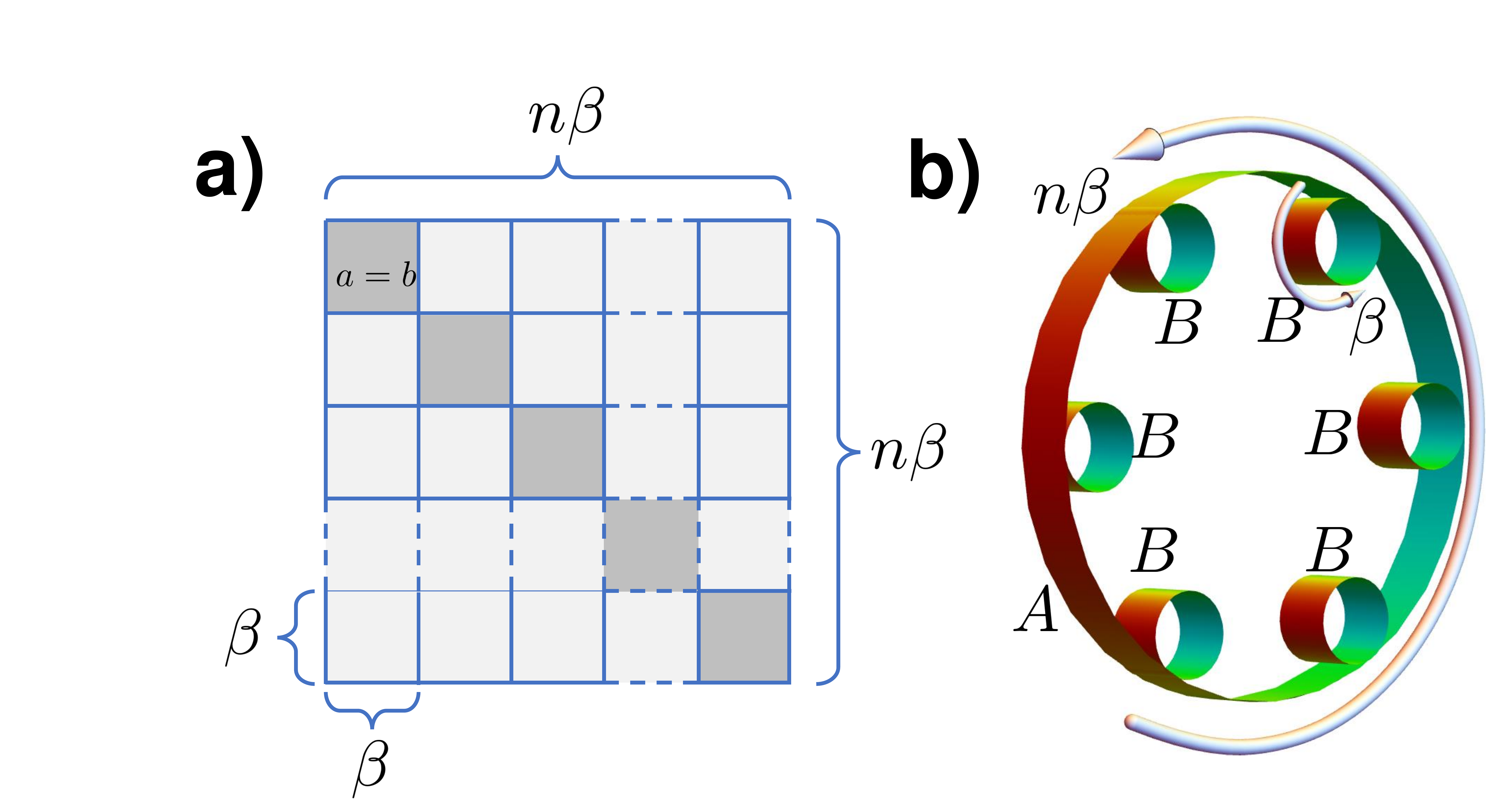}
\caption{illustration of the boundary conditions of $G$ and $g$. a). The argument region of $G(\tau,\tau')$ and $g(\tau,\tau')$. The horizontal axis and the vertical axis are for $\tau$ and $\tau'$, respectively. b). Antiperiodical boundary conditions along $\tau$ and $\tau'$ [Eq.~\eqref{BCs}] for subsystems $A$ and $B$. The $B$ cycles have periodicity $\beta$ and carry the replica index $a, b$ ranging from 1 to $n$, while the $A$ cycle has antiperiodicity $n\beta$ and has only one copy.}\label{IllusFig}
\end{figure}

\subsection{Replica trick method}
\label{sec:replica}
The local observables, correlation functions, and thermodynamics of the SYK$q$ models are exactly obtained in the large $N$ limit using the path integral formalism\cite{Kitaev15talk, Madacena16Remarks}.
It is natural to ask whether the EE of the SYK$q$ models may also be found exactly.  Using the standard replica method and path integral formalism, one may write
\begin{equation}
\overline{S_A} 
= \lim\limits_{n\rightarrow 1} \frac{ \overline{\ln \mathrm{Tr}_A[\rho_A^n]}}{1-n}
=\lim\limits_{n\rightarrow 1} \frac{ \overline{\ln (Z_n/Z^n)}}{1-n},
\end{equation}
where $Z_n$ is the partition function involving $n$ copies of $B$ coupled with A as shown in Fig. \ref{IllusFig}b). The $n\rightarrow 1$ limit above is used to obtain the logarithm in the definition of the von Neumann entropy.  Notice that for a disordered system, to compute $\overline{\ln Z}$ we need in principle to perform a {\em second} replica trick. Luckily,  the SYK$q$ in the large N limit is always in a replica symmetric phase and therefore we have $\overline{\ln (Z_n/Z^n)}=\ln\overline{Z_n}-n\ln\overline{Z}$.

After disorder average and a saddle point approximation valid at large $N$, we have
$\overline{Z}_n = \exp(-N S_{n,\text{cl}})$. The zeroth order physics is determined entirely by the saddle point equation (see Appendix \ref{sec: SaddleSYK4} for the derivation):
\begin{subequations}\label{EOM}
\begin{eqnarray}
\mathcal{J}^2[\lambda G^{ab}(\tau,\tau')+(1-\lambda)g^{ab}(\tau,\tau')]^3&=&\Sigma^{ab}(\tau,\tau'),\label{EoM1}\\
\frac{1}{\partial^A_\tau \delta(\tau-\tau')\delta^{ab}-\Sigma^{ab}(\tau,\tau')} &=& G^{ab}(\tau,\tau'),\label{EoM3}\\
\frac{1}{\partial^B_\tau \delta(\tau-\tau')\delta^{ab}-\Sigma^{ab}(\tau,\tau')} &=& g^{ab}(\tau,\tau'),\label{EoM4}
\end{eqnarray}
\end{subequations}
where $1\leq a,b\leq n$ are the replica indices, $G$ and $g$ are the Green's functions, respectively, for the $A$ and $B$ subsystem. $G$ and $g$ have different boundary conditions
\begin{equation}\label{BCs}
\begin{aligned}
G^{ab}(\beta,\tau') = G^{(a+1)b}(0^-,\tau'),&\quad G^{nb}(\beta,\tau') = -G^{0b}(0^-,\tau'),\\
G^{ab}(\tau,\beta) = G^{a(b+1)}(\tau,0^-),&\quad G^{an}(\tau,\beta) = -G^{a0}(\tau,0^-),\\
g^{ab}(\beta,\tau') = -g^{ab}(0^-,\tau'),&\quad g^{ab}(\tau,\beta) = -g^{ab}(\tau,0^-),
\end{aligned}
\end{equation}
where $a=1,\ldots,n-1$,$b=1,\ldots,n$ for the first line, $a=1,\ldots,n$, $b=1,\ldots,n-1$ for the second line and $a,b=1,\ldots,n$ for the third line. The differential operators $\partial^{A/B}_\tau$ are different for $A$ and $B$ which ensure the continuity of the equations (see Appendix \ref{sec: SaddleSYK4} for explanation). These boundary conditions are illustrated in Fig.~\ref{IllusFig}. Due to the diffferent boundary conditions for $G$ and $g$, obtaining a general and explicit form for $G$ and $g$ turns out to be hard (Similar issue arises in entanglement dynamics in a SYK chain model in Ref.\ \onlinecite{gu2017spread}). However, the EE in the scaling limit $\lambda\rightarrow 0$ does not require solving Eqs.~\eqref{EOM} and can be readily obtained by counting degrees of freedom. In this limit, only the diagonal term in $g^{ab}$ is nonzero and is the same as that in $\overline{Z}=\exp(-NS_{\text{cl}})$. Moreover, $G^{ab} \simeq g^{ab}\delta_{ab}$ to first order of $\lambda$.  As a result, all terms in $\ln\overline{Z}_n$ cancel those in $n\ln \overline{Z}$, except for the term $\ln \mathrm{det}(\partial^A_\tau)$ that counts the degrees of freedom of subsystem $A$. In $\overline{Z}_n$, there is only one $A$ subsystem with time antiperiodicity $n \beta$, while there are $n$ copies in $\overline{Z}^n$. Therefore we have
\begin{equation}
\frac{\overline{Z}_n}{\overline{Z}^n} =  \exp \left[-\frac{n-1}{2} \lambda\ln \mathrm{det} (\partial^A_\tau)\right] = \exp \left[-(n-1) \lambda \left(\begin{smallmatrix}N\\ \kappa N\end{smallmatrix}\right)\right].
\end{equation}
This result is independent of $q$ and leads to Eq. \eqref{SAlambda0}.  This shows the super-universality of maximal entanglement for small subsystems for all $q$.

For $\lambda$ around $1/2$, solving Eqs.~\eqref{EOM} is difficult. Notably, one must impose anti-periodicity in time $T=n\beta$ on the $A$ subsystem, which couples to all $n$ $B$-replicas. This coupling introduces an effective interaction between different $B$-replicas, which leads to nonvanishing $g^{ab}$ even for $a\neq b$. In principle Eqs.~\eqref{EOM} can be solved numerically but appropriate methods have to be developed and analytic continuation to $n=1$ needs to be understood. We leave this to future work.

\section{Discussion and Outlook}
\label{sec:conclusion}
In this paper, we studied the entanglement aspects of the Sachdev-Ye-Kitaev (SYK) model at $q=2$ and $4$. For the free model SYK$2$, we showed that the reduced correlation matrix belongs to the $\beta$-Jacobi ensemble at $\beta=2$, and obtained from this an analytic expression for the EE of SYK$2$. We further demonstrated that the ground state EE of the SYK$4$ model shows a derivation from the Page value which persists to the thermodynamic limit, indicating that SYK$4$ is not maximally entangled.   Furthermore, we explored the EE for the SYK$4$+SYK$2$ model, and found that there are two distinct regimes as we vary energy: for  low energy states, the EE is dominated by the SYK2 term, while for  highly excited states, the physics is dominated by the SYK4 term and the EE is the same as the Page value.

We were unable to obtain a full analytical expression for the EE of the SYK4 model for an arbitrary bipartition. This calculation requires coupling of different replicas and it would be very interesting to develop a field theoretic approach to this problem in the large $N$ limit, even if partly numerical.\\

{\it Note added.} 
Shortly after the submission of this draft to arXiv, another paper \cite{2017arXiv170909160H} gave the analytical estimation of the EE of the SYK4 model at subsystem/system=1/2, which is consistent with our numerical result at this ratio.

\begin{acknowledgements}  
We thank A.~Ludwig, Z.~Bi, L.~Vidmarand, V.~Rosenhaus and P.~Lu for useful discussions. X.C. was supported by a postdoctoral fellowship from the the Gordon and Betty Moore Foundation, under the EPiQS initiative, Grant GBMF4304, at the Kavli Institute for Theoretical Physics. We acknowledge support from the Center for Scientific Computing from the CNSI, MRL: an NSF MRSEC (DMR-1121053).  L.B. and C.L. were supported by the NSF Materials Theory program, grant number DMR1506199.
\end{acknowledgements}

\appendix
\section{$\beta-$Jacobi Ensemble and its eigenvalue scaling laws}\label{sec: Jacobi}

Suppose matrices $A$ and $B$ are of size $m_1\times m$ and $m_2\times m$, respectively, both with independently and identically distributed Gaussian entries over either the real, complex or the quaternion field ($\beta=1,2,4$ respectively). Then by $\beta$-Jacobi (or MANOVA) ensembles we mean the eigenvalue distribution of 
\begin{equation}\label{MANOVA}
 \frac{ A^\dag A}{A^\dag A+B^\dag B}.
\end{equation}
The joint eigenvalue distribution is
\begin{equation}\label{joint}
\begin{aligned}
f(\varepsilon_1,\cdots, \varepsilon_m) &= C(m_1,m_2,m) \prod\limits_{1\leq i < j \leq m}|\varepsilon_i-\varepsilon_j|^\beta \\
&\cdot \prod\limits_{i=1}^m \varepsilon_i^{\frac{\beta}{2}(m_1-m+1)-1}(1-\varepsilon_i)^{\frac{\beta}{2}(m_2-m+1)-1}.
\end{aligned}
\end{equation}

In the limit $m_1,m_2,m\rightarrow \infty$ while keeping the ratios
$$\lambda = \frac{m}{m_1+m_2},\qquad \kappa = \frac{m_1}{m_1+m_2}$$
finite, the eigenvalue distribution satisfies the Wachter law [Eq.~(2) in the main text].

The EE of the complex SYK2 model is calculated via the reduced correlation function $C_A$. The eigenvalues of $C_A$ belong to the $\beta$-Jacobi ensemble, which in the large $N$ limit satisfy the Wachter law. This is because the single-particle Hamiltonian of complex SYK2 is a GUE, so the eigenvector matrix $U$ is uniformly distributed in the space $U(N)$ according to the Haar measure. The truncated correlation matrix $C_A=V^\dag V$ therefore satisfies Eq. \eqref{MANOVA}, provided we set $A=V$, $B=W$, $m_1=n$ and $m_2 = N-n$, where
$$ U  =  \left(\begin{array}{cc} V_{n\times m} & V'_{n\times( N-m)}\\ W_{(N-n)\times m} & W'_{(N-n)\times (N-m)}\end{array}\right)^T.$$

The derivation from Eq.~\eqref{MANOVA} to Eq.~\eqref{joint} is achieved by diagonalizing \eqref{MANOVA}, which induces a basis transformation in the space of matrices; the ``level repulsion'' term $\prod|\varepsilon_i-\varepsilon_j|^\beta$ comes from the Jacobian of the measure during transformation. Below we sketch the derivation from Eq.~\eqref{joint} to Eq.~(2) in the main text, known as the \textsl{Coulomb gas approach}, and for simplicity we only discuss $\beta=2$. For a complete proof, see Ref. \cite{Spectral12proof}. 

The Coulomb gas approach expresses the joint distribution as a path integral. In the limit $m_1,m_2,m\rightarrow \infty$ (keeping $\lambda$ and $\kappa$ finite) we have
$$f(\varepsilon_1,\cdots,\varepsilon_m) \rightarrow \frac{1}{Z} e^{-NS[\varepsilon]},$$
where $N = m_1+m_2$, and
$$S = -\iint d xd y \ln |\varepsilon(x)-\varepsilon(y)|- \int dx (a \ln \varepsilon + b \ln (1-\varepsilon)),$$
where $a = \kappa - \lambda$, $b = 1-\kappa - \lambda$. Due to the presence of the large $N$ in front of the action, the leading order behavior is entirely determined by the saddle point equation
$$ \int \frac{d y}{|\varepsilon(x) -\varepsilon (y)|}+ \frac{a}{\varepsilon(x)}-\frac{b}{1-\varepsilon(x)}=0.$$
Under appropriate boundary conditions, this equation has unique physical solution, Eq.~(2) in the main text. 

\begin{figure}
\centering
\includegraphics[width=0.5\textwidth]{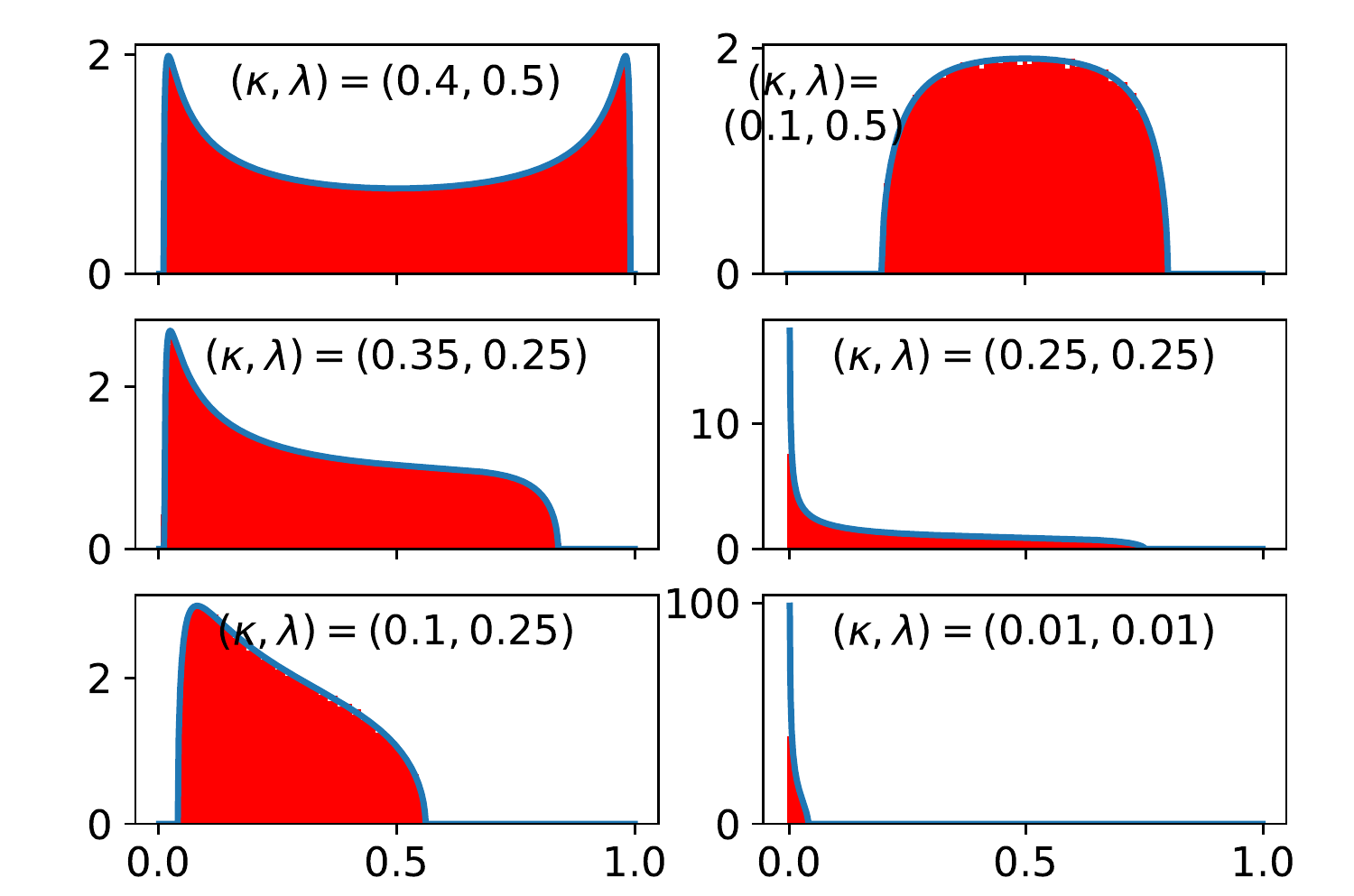}
\caption{Wachter law for more $(\kappa,\lambda)$ values.}
\label{App:entanglement_spectrum}
\end{figure}

The Wachter distribution $f(x,\kappa,\lambda)$ for some generic values of $(\kappa,\lambda)$ are shown in Fig.~\ref{App:entanglement_spectrum}. 

\section{Calculating Entanglement Entropy of Majorana SYK2 Using Correlation Functions}\label{sec:Maj}

The Majorana SYK2 model is
\begin{equation}
H_{2,\chi} = i\sum_{i,j =1}^N J_{ij} \chi_i \chi_j = \chi^T J \chi,
\end{equation}
where $\chi_i, i=1,...,N$ are Majorana fermion operators with $\chi^\dag = \chi$, and $\{ \chi_i, \chi_j \} = 2\delta_{ij}$. Here $N$ is even and we set $N=2M$.

To ensure hermiticity, $J$ is antisymmetric with dimension $2M\times 2M$. $J$ can then be ``diagonalized'' to the block form below by a $2M\times 2M$ orthogonal matrix $O$:
$$J = O\Sigma O^T,$$
where
\begin{equation}\label{block}
\Sigma = \left(\begin{array}{cccccccc}
0&&&&&&&\\
&\ddots&&&&&&\\
&&0&&&&&\\
&&&0&\lambda_1&&&\\
&&&-\lambda_1&0&&&\\
&&&&&\ddots&&\\
&&&&&&0&\lambda_r\\
&&&&&&-\lambda_r&0
\end{array}\right),
\end{equation}
where there are $2(M-r)$ zeros on the diagonal block, and
$$0<\lambda_1\leq \lambda_2\leq \cdots \leq \lambda_r.$$
Now let us denote $\gamma = O^Tc = \left(\gamma^1_{r+1},\gamma^2_{r+1},\cdots, \gamma^1_M,\gamma^2_M,\gamma^1_1,\gamma^2_1,\cdots, \gamma^1_r,\gamma^2_r\right)^T$, where $\gamma^\alpha_j$ ($j=1,...,M$, $\alpha=1,2$) denote the $2M$ rotated Majorana operators which satisfy $\left(\gamma^\alpha_i\right)^\dag=\gamma^\alpha_i$ and $\{\gamma^\alpha_i,\gamma^\beta_j\} = 2\delta_{\alpha\beta}\delta_{ij}$. We have
\begin{equation}\label{h1}
H = i \chi^T J\chi = i \chi^T O \Sigma O^T \chi = i \gamma^T \Sigma  \gamma
= \sum_{i=1}^r 2i \lambda_i \gamma^1_i\gamma^2_i.
\end{equation}

We further denote
$\gamma^1_j = f_j + f^\dag_j$ and $\gamma^2_j = \frac{1}{i} (f_j-f_j^\dag)$, $j=1,...,M$, where $f,f^\dag$'s are complex fermion operators with standard anticommutation relations, and
\begin{equation}\label{basi}
\gamma^1_j\gamma^2_j = \frac{1}{i}(f_j+f_j^\dag)(f_j-f_j^\dag) = \frac{1}{i}(2f_j^\dag f_j - 1),
\end{equation}
then $H_{2,\chi}$ is diagonalized by $f_i$'s
\begin{equation}
H_{2,\chi} = \sum\limits_{i=1}^r 2 \lambda_i (2 f^\dag_i f_i -1),
\end{equation}
where states labeled by $i=1,...,r$ have finite excitation energy $4 \lambda_i$, and $i=r+1,...,M$ have zero excitation energy. 

For the rest we will assume that $r=M$, i.e. the Hamiltonian $H_{2,\chi}$ has no zero modes. This is justified for the SYK2 model since a random matrix $J$ has almost zero possibility in producing zero eigenvalues. The ground state of SYK2 is then $|\Omega\rangle = |0\rangle$, where $|0\rangle$ is the state that is annihilated by all $f$: $f_i|0\rangle =0$ for $i=1,...,M$. 
The ground state expectation value is therefore $\langle \gamma^1_j \gamma^2_j\rangle = -\langle \gamma^2_j \gamma^1_j\rangle  = \frac{1}{i} \langle (2 f^\dag_j f_j - 1)\rangle
=  i$, and 
$$\langle \gamma \gamma^T\rangle 
=
\left(\begin{array}{ccccc}
1&i&&&\\
-i&1&&&\\
&&\ddots&&\\
&&&1&i\\
&&&-i&1
\end{array}\right)\equiv I+i K,$$
where we have defined 
$$K = -I_{M\times M}\otimes \left(\begin{array}{cc} 0& 1\\-1 &0\end{array}\right).$$

Then, the correlation function matrix becomes
$$C = \langle \chi \chi^T\rangle = \langle O\gamma \gamma^T O^T\rangle
=O\langle \gamma \gamma^T\rangle O^T = I + iO K O^T, $$
and the reduced correlation function matrix is (for subsystem $A$ with size $|A|=2m$)
$$C_A = I_{2m\times 2m} + iO_{2m\times 2M} K_{2M\times 2M} O^T_{2m\times 2M}.$$
After diagonalizing $C_A$ we get $C_A = I_{2m\times 2m} + iY^\dag E Y$, where
$$E = \mathrm{Diag}\left( \varepsilon_1,\cdots,\varepsilon_m\right)\otimes \left(\begin{array}{cc} 0& 1\\-1 &0\end{array}\right),$$
and we have \cite{Peschel02Reduced}
\begin{equation}\label{EE1101}
S_A = -\sum\limits_{i=1}^m \frac{1-\varepsilon_i}{2} \ln \frac{1-\varepsilon_i}{2} + \frac{1+\varepsilon_i}{2} \ln \frac{1+\varepsilon_i}{2}.
\end{equation}

To prove Eq.~\eqref{EE1101}, we assume that the reduced density matrix can be written in terms of an effective Hamiltonian $H_A= \chi^T J_A \chi$, where $J_A$ is a $2m\times 2m$ matrix, by
$$\rho_A = \frac{1}{Z}e^{-H_A},\qquad Z = \mathrm{Tr}_A[e^{-H_A}]. $$
Suppose $H_A$ is diagonalized in some Majorana basis $\xi$ and some complex basis $\psi$ (following the derivation for diagonalizing $H_{2,\chi}$)
$$H_A = \chi^T J_A \chi =\sum\limits_{i=1}^{m} 2 i h_i \xi^1_i \xi^2_i=\sum\limits_{i=1}^m 2 h_i (2 \psi^\dag_i \psi -1),$$
then $\rho_A$, as a $2^m\times 2^m$ matrix that acts on the many-body Hilbert space of $A$, has eigenvalues
$$ \widetilde{\varepsilon}_i = \frac{e^{-2h_i}}{e^{2h_i}+e^{-2h_i}}, \quad i=1,2,...,m,$$
and finally, by definition, the EE is
\begin{equation}\label{EE11001}
\begin{aligned}
S_A &= -\mathrm{Tr} \rho_A \ln \rho_A
= -\sum\limits_{D \in \mathcal{D}} \prod_{\alpha \in D} \prod_{\beta \in A\backslash D}\widetilde{\varepsilon}_\alpha  (1-\widetilde{\varepsilon}_\beta) \ln  \widetilde{\varepsilon}_\alpha  (1-\widetilde{\varepsilon}_\beta)\\
&=-\sum\limits_{i=1}^m \widetilde{\varepsilon}_i \ln \widetilde{\varepsilon}_i -\sum\limits_{i=1}^m (1-\widetilde{\varepsilon}_i) \ln(1-\widetilde{\varepsilon}_i). 
\end{aligned}
\end{equation}

To connect $\rho_A$ with the truncated correlation function matrix $C_A$, note that $\langle \xi \xi^T\rangle$ is a block diagonal matrix, $\langle \xi^1_j\xi^2_j\rangle = -\langle \xi^2_j \xi^1_j \rangle = \frac{1}{i}(2 \langle \psi^\dag_j \psi_j\rangle -1)$ and 
\begin{equation}
\begin{aligned}
\langle \psi^\dag_j \psi_j\rangle&= \frac{\sum\limits_{\{ n\}_{i=1}^m =0,1} \langle \{n\}_{i=1}^m |e^{- \sum\limits_{i=1}^m 2h_i (2 \psi^\dag_i \psi_i -1)} \psi^\dag_j \psi^\dag_j|\{n\}_{i=1}^m \rangle}{\sum\limits_{\{ n\}_{i=1}^m =0,1} \langle \{n\}_{i=1}^m |e^{- \sum\limits_{i=1}^m 2h_i (2 \psi^\dag_i \psi_i -1)} |\{n\}_{i=1}^m \rangle}\\
&=\frac{\sum\limits_{n_j = 0,1}\langle n_j | e^{-2 h_j(2 \psi^\dag_j \psi_j-1)} \psi^\dag_j \psi_j|n_j\rangle}{e^{2 h_i}+e^{-2 h_i}}\\
&=\widetilde{\varepsilon}_i.
\end{aligned}
\end{equation}
These lead to 
$$C_A =  I -  iY^\dag_A E_A Y_A,$$
where
$$E_A = \mathrm{Diag}(2\widetilde{\varepsilon}_1-1,\cdots 2\widetilde{\varepsilon}_m-1)\otimes \left(\begin{array}{cc} 0& 1\\-1 &0\end{array}\right),$$
thus the eigenvalues of $C_A$ are $\varepsilon_i = \pm (2 \widetilde{\varepsilon}_i-1)$, $i=1,...,M$. Plug these into Eq.~\eqref{EE11001} we get Eq.~\eqref{EE1101}.

\onecolumngrid

\section{Path integral formalism for the Entanglement Entropy of SYK4}\label{sec: SaddleSYK4}

The path integral formalism helps find the saddle point equations for the SYK models at leading order of $N$. In this Appendix, we sketch the derivation for SYK4 but the generalization to all $q$ is straightforward.
We introduce $n$ replicas to SYK$4$, with index $a=1,...,n$. Notice that
\begin{equation}
\begin{aligned}
Z_n\left(\{J_{ijkl}\}\right)&=\mathrm{Tr}_A\rho^n_A \\
&= 
\mathrm{Tr}_A\left(\mathrm{Tr}_B \rho\right)^n\\
&=\int d\psi_A \langle \psi_A | \left( \int \left[d\psi^a_B\right]_{a=1}^n \langle \psi^1_B| \rho|\psi^1_B\rangle \cdots
\langle \psi^n_B| \rho|\psi^n_B\rangle \right) |\psi_A\rangle\\
&=
\int \left[\mathcal{D}\chi^a_i\right]_{a,i} \exp\left[\int^\beta_0 d\tau \sum\limits_{a=1}^n  \left(-\frac{1}{2}\chi^a_i\partial^a_\tau \chi^a_i- \sum\limits_{1\leq i<j<k<l\leq N} J_{ijkl} \chi^a_i(\tau) \chi^a_j(\tau)\chi^a_k(\tau) \chi^a_l(\tau)\right)\right],
\end{aligned}
\end{equation}
where we use $|\psi_B^a\rangle$, $a=1,...,N$, to denote a complete basis for the Hilbert space of subsystem $B$, and $|\psi_A\rangle$ a complete basis for the Hilbert space of subsystem $A$. In the last line $\chi^a_i$ become real Grassmann fields with a continuous index $\tau \in [0,\beta)$ and a discrete replica index $a=1,...,N$. Because of the different trace rules in $A$ and $B$, the boundary conditions for $\chi$ are
$$\chi^a_i(\beta)=\chi^{a+1}_i(0^-)\text{ and } \chi^n_i(\beta)=-\chi^{1}_i(0^-)\quad\text{for } i\in A, \qquad \chi^b_i(\beta) = -\chi^b_i(0^-)\quad \text{for } i\in B,$$
where $a=1,\ldots,n-1$ and $b=1,\ldots,n$. The operators $\partial_\tau$ for $A$ and $B$, accordingly, also differ, respecting the boundary condtions for $A$ and $B$, respectively, which ensures they are continuous operators. To remind these differences, we also label $\partial_\tau$ with replica index $a$, where $\partial_\tau^a=\partial_\tau^A$ ($\partial_\tau^b=\partial_\tau^B$) are the same for all $a \in A$ ($b\in B$).

After averaging over $J_{ijkl}$ we have
\begin{equation}\label{Znn}
\begin{aligned}
\overline{Z}_n&=\int \left(\prod_{1\leq i<j<k<l\leq N}dJ_{ijkl} \frac{1}{\sqrt{2\pi \sigma}} e^{-\frac{J^2_{ijkl}}{2\sigma^2}} \right) 
Z_n\left(\{J_{ijkl}\}\right)\\
&=\int \left[\mathcal{D}\chi^a_i\right]_{a,i} \exp\left[-\frac{1}{2}\int^\beta_0 d\tau \sum\limits_{a=1}^n \chi^a_i\partial^a_\tau \chi^a_i+\frac{\mathcal{J}^2}{8N^3}  \int^\beta_0 d\tau d\tau' \sum\limits_{a,b=1}^n \left(\sum\limits_{i=1}^N\chi^a_i(\tau) 
\chi^b_i(\tau')\right)^4
\right],
\end{aligned}
\end{equation}
in which the $\chi$ fields for different sites are decoupled. Suppose $A$ has $P$ particles and $B$ has $N-P$ particles. We introduce Green's functions $G$ and $g$
and self-energies $\Sigma$ and $\sigma$ such that
\begin{equation}
\begin{aligned}
1 &= \int \left[\mathcal{D} G^{ab}\mathcal{D} \Sigma^{ab}\right]_{a,b}\exp\left[-\frac{1}{2}\int^\beta_0d\tau d\tau' \sum\limits_{a,b=1}^n\Sigma^{ab}(\tau,\tau')\left(PG^{ab}(\tau,\tau')-\sum_{i\in A} \chi^a_i(\tau)\chi^b_i(\tau')\right)\right],\\
1 &= \int \left[\mathcal{D} g^{ab}\mathcal{D} \sigma^{ab}\right]_{a,b}\exp\left[-\frac{1}{2}\int^\beta_0d\tau d\tau' \sum\limits_{a,b=1}^n\sigma^{ab}(\tau,\tau')
\left((N-P)g^{ab}(\tau,\tau')-\sum_{i\in B} \chi^a_i(\tau)\chi^b_i(\tau')\right)\right],
\end{aligned}
\end{equation}
plug them into Eq.~\eqref{Znn} and then conduct the integrals over the Majorana fields $\chi_i$, we get
\begin{equation}
\begin{aligned}
\overline{Z}_n
&=\int
\left[\mathcal{D} G^{ab}\mathcal{D} \Sigma^{ab}
      \mathcal{D} g^{ab}\mathcal{D} \sigma^{ab}\right]_{a,b}\exp\left[P\ln\mathrm{Pf} \left(\partial^A_\tau\delta(\tau-\tau')\delta^{ab} - \Sigma^{ab}(\tau,\tau')\right)
+
(N-P)\ln\mathrm{Pf} \left(\partial^B_\tau\delta(\tau-\tau')\delta^{ab} - \sigma^{ab}(\tau,\tau')\right)\right]\\
 &\times
\exp\left[\int^\beta_0d\tau d\tau' \sum_{a,b=1}^n
\left(-\frac{P}{2}\Sigma^{ab}(\tau,\tau')G^{ab}(\tau,\tau')-\frac{N-P}{2}\sigma^{ab}(\tau,\tau')g^{ab}(\tau,\tau')+\frac{\mathcal{J}^2}{8N^3}\left[PG^{ab}(\tau,\tau')+(N-P)g^{ab}(\tau,\tau')\right]^4\right)\right],
\end{aligned}
\end{equation}
which gives back Pfaffians of matrices $\partial^A_\tau\delta(\tau-\tau')\delta^{ab} - \Sigma^{ab}(\tau,\tau')$ and $\partial^B_\tau\delta(\tau-\tau')\delta^{ab} - \sigma^{ab}(\tau,\tau')$, with index pairs $(a,\tau)$ as rows and $(b,\tau')$ as columns. Also note that
$G^{ab}(\tau,\tau')$ and $g^{ab}(\tau,\tau')$ respect the boundary conditions of the $\chi's$ in $A$ and $B$, which gives Eq.~(8) in the main text.
Now define $P = \lambda N$, thus $N-P = (1-\lambda)N$, $\lambda\in [0,1]$ and we have a large parameter $N$ that controls the behavior of $\overline{Z}_n$. In the thermodynamic limit $N\rightarrow \infty$, we have
\begin{equation}
\overline{Z}_n = \exp(-NS_{\text{cl}}),
\end{equation}
i.e. $\overline{Z}_n$ is dominated entirely by the classical action $S_{\text{cl}}$. $S_{\text{cl}}$ is obtained by plugging in the solution of saddle point equations for $G^{ab}(\tau,\tau')$, $g^{ab}(\tau,\tau')$, $\Sigma^{ab}(\tau,\tau')$ and $\sigma^{ab}(\tau,\tau')$. The saddle point equations for $G^{ab}(\tau,\tau')$, $g^{ab}(\tau,\tau')$, $\Sigma^{ab}(\tau,\tau')$ and $\sigma^{ab}(\tau,\tau')$ in $\overline{Z}_n$ are
\begin{subequations}\label{EoM}
\begin{eqnarray}
\lambda\Sigma^{ab}(\tau,\tau')-\lambda \mathcal{J}^2[\lambda G^{ab}(\tau,\tau')+(1-\lambda)g^{ab}(\tau,\tau')]^3&=&0,\label{EoM1}\\
(1-\lambda)\sigma^{ab}(\tau,\tau')-(1-\lambda)\mathcal{J}^2[\lambda G^{ab}(\tau,\tau')+(1-\lambda)g^{ab}(\tau,\tau')]^3&=&0,\label{EoM2}\\
\frac{1}{\partial^A_\tau \delta(\tau-\tau')\delta^{ab}-\Sigma^{ab}(\tau,\tau')} &=& G^{ab}(\tau,\tau'),\label{EoM3}\\
\frac{1}{\partial^B_\tau \delta(\tau-\tau')\delta^{ab}-\sigma^{ab}(\tau,\tau')} &=& g^{ab}(\tau,\tau').\label{EoM4}
\end{eqnarray}
\end{subequations}
Solving Eqs.~\eqref{EoM} is a hard task due to the different boundary conditions of $A$ and $B$, and as explained in the main text, the solutions are no longer in the replica symmetric phase $g^{ab}=g\delta^{ab}$. However using Eq.~\eqref{EoM1} and Eq.~\eqref{EoM2} we do get
\begin{equation}\label{Sequaltos}
\Sigma^{ab}(\tau,\tau') = \sigma^{ab}(\tau,\tau'),\qquad a,b=1,...,n,\quad  \tau,\tau' \in [0,\beta).
\end{equation}
We leave solving the complete version of Eqs.~\eqref{EoM} to future work.

\bibliography{EE_SYK}


\end{document}